\documentclass[aps,prb,preprint,groupedaddress,showpacs,amsmath,amssymb,nofootinbib]{revtex4-2}

\usepackage{graphicx}% Include figure files
\usepackage{dcolumn}% Align table columns on decimal point
\usepackage{bm}% bold math
\usepackage{amsmath}
\usepackage{hyperref}

\begin{document}

\title{Effects of physical and chemical pressure on charge density wave transitions in LaAg$_{1-x}$Au$_x$Sb$_2$ single crystals}

\author{Li Xiang $^{1,}$\footnote{Current address: National High Magnetic Field Laboratory, Florida State University, Tallahassee, Florida 32310, USA}, Dominic H. Ryan $^{1,2}$, Paul C. Canfield  $^{1}$and Sergey L. Bud'ko $^{1}$}

\affiliation{$^{1}$Ames National Laboratory, US DOE, and Department of Physics and Astronomy, Iowa State University, Ames, Iowa 50011, USA}
\affiliation{$^{2}$Physics Department and Centre for the Physics of Materials, McGill University, 3600 University Street, Montreal, Quebec, Canada H3A 2T8}

\date{\today}

\begin{abstract}

The structural characterization and electrical transport measurements at ambient and applied pressures of the compounds of the La(Ag$_{1-x}$Au$_x$)Sb$_2$ family are presented. Up to two charge density wave (CDW) transitions could be detected upon cooling from room temperature and an equivalence of the effects of chemical and physical pressure on the CDW ordering temperatures was observed with the unit cell volume being a salient structural parameter.  As such La(Ag$_{1-x}$Au$_x$)Sb$_2$ is a rare example of a non-cubic system that exhibits good agreement between the effects of applied, physical, pressure and changes in unit cell volume from steric changes induced by isovalent substitution. Additionally, for La(Ag$_{0.54}$Au$_{0.46}$)Sb$_2$ anomalies in low temperature electrical transport were observed in the pressure range where the lower charge density wave is completely suppressed.

\end{abstract}

%\pacs{}
%\keywords{}
\maketitle

%%%%%%%%%%%%%%%%%%%%%%%%%%%%%%%%%%%%%%%%%%

% The order of the section titles is different for some journals. Please refer to the "Instructions for Authors” on the journal homepage.

\section{Introduction}

Physical (hydrostatic) pressure and chemical substitution are two common ways to tune the physical properties of materials. Whereas  hydrostatic pressure is considered to be a clean  parameter that does not introduce additional disorder, as well as changes in band filling in many cases, the experimental techniques available under pressure are more limited. Chemical substitution necessarily involves some additional disorder. In the case of aliovalent substitutions the corresponding electron- or hole- doping effects are often dominant. For isovalent substitution the primary effect is thought to be steric and the comparison with physical pressure can be more relevant. Whereas such isovalent substitutions can be referred to as "chemical pressure" differences in how pressure and substitution affect a compound, especially a non-cubic one, can be greater than similarities in some cases. It is of particular importance when there is a desire to stabilize a particular high pressure phase/state (like high temperature superconductivity \cite{wan21a}) using chemical and/or physical pressure. Additionally, observation of an apparent equivalence \cite{kli10a,pag10a,kim11a} or non-equivalence \cite{fer91a} of chemical and physical pressure can help in understanding of structure - property relations and in recognizing relevant structural motifs.

The  members of the family of compounds chosen for this study, LaAg$_{1-x}$Au$_x$Sb$_2$, demonstrate charge density wave (CDW) transitions, or spontaneous superstructures formed by electrons. \cite{mon12a} Decades ago Peierls showed the instability of a (one-dimensional) metal interacting with the lattice towards a lattice distortion and the opening of a gap in the electronic spectrum. \cite{pei55a} This concept is often applied to CDW formation in low-dimensional materials, although alternatives are widely discussed. \cite{joh08a,eit13a,zhu15a} Studies of CDW phenomena in solids and competition of CDW with other collective phenomena remain one of the active subfields of quantum materials research.

In this work we study effects of pressure on single crystals of the selected members of the LaAg$_{1-x}$Au$_x$Sb$_2$ family which is compared to the chemical pressure implemented via Ag $\leftrightarrow$ Au substitution. The end-compounds, LaAgSb$_2$ and LaAuSb$_2$  were first synthesized almost three decades ago \cite{bry95a,sol94a} and were reported to crystallize in the same tetragonal ZrCuSi$_2$-type structure ($P4/nmm$, No. 129). CDW transitions were observed in electrical transport at $\sim 210$~K and $\sim 100$~K  for  LaAgSb$_2$ and LaAuSb$_2$ respectively. \cite{mye99a, seo12a} Synchrotron X-ray scattering study \cite{son03a} and further thermodynamic measurements \cite{bud08a} identified second, lower temperature, CDW transition at $\sim 185$~K in LaAgSb$_2$. Similarly, in addition to $T_{CDW1} \approx 110$~K, a second CDW transition at $T_{CDW2} \sim 90$~K was detected by electrical transport measurements in near stoichiometric LaAuSb$_2$. \cite{xia20a} The synthesis and evolution of the higher temperature, $T_{CDW1}$ in the  LaAg$_{1-x}$Au$_x$Sb$_2$ series was reported in Ref. \cite{mas14a} but without any measurements of $T_{CDW2}$ or companion applied pressure studies. 

For LaAgSb$_2$ pressure reportedly suppressed $T_{CDW1}$ \cite{bud06a,tor07a,aki21a,zha21a,aki22a} as well as $T_{CDW2}$ \cite{aki21a,aki22a} with the results being consistent in majority of publications. \cite{bud06a,tor07a,aki21a} Qualitatively similar behavior under pressure was also observed for LaAuSb$_2$. \cite{xia20a,duf20a,lin21a} Additionally, in LaAgSb$_2$ and LaAuSb$_2$, low temperature superconductivity was discovered and studied under pressure \cite{aki22a,duf20a}. It has to be noted that the  exact Au stoichiometry in  LaAu$_x$Sb$_2$ depends on details of the synthesis and affects both  the ambient pressure values of $T_{CDW1}$ and $T_{CDW2}$ and their pressure derivatives. \cite{xia20a}

Earlier comparison of the effects of  pressure  and chemical substitution in this family of materials \cite{bud06a} was  based on a study of the La$_{1-x}$$R_x$AgSb$_2$ series ($R$ = Y, Ce, Nd, Gd) and significant contribution of disorder prevailing in the case of substitution was found thus resulting in significant difference between physical and chemical pressure. In this work we address the same question in the different, transition metal site, substitution series.

%%%%%%%%%%%%%%%%%%%%%%%%%%%%%%%%%%%%%%%%%%
\section{Materials and Methods}

Single crystals of LaAg$_{1-x}$Au$_x$Sb$_2$  were grown from an antimony-rich self-flux following the method described in Refs. \cite{mye99a,zha16a,xia20a}. Pure elements  were loaded into  an alumina Canfield crucible set \cite{can16a} which was  placed into an amorphous silica tube and sealed in partial atmosphere of argon. The sealed tubes were heated to 1050~$^{\circ}$C over  10~hours, held for 8~hours, then cooled to 800~$^{\circ}$C over a period of 10~hours prior to starting the crystal growth. Crystal growth occurred during the 100~hour cooling from 800~$^{\circ}$C to 670~$^{\circ}$C, after which the excess flux was decanted with the aid of a centrifuge. 

In this work, crystals of LaAg$_{1-x}$Au$_x$Sb$_2$ with nominal compositions $x = 0, 0.25, 0.5, 0.75$ were grown with the initial La :  T : Sb (T = Ag$_{1-x}$Au$_x$) growth compositions:  1 : 2 : 20 (T2).  To investigate whether reported Au deficiency \cite{xia20a,seo12a,mas14a} is relevant and can be tuned for the intermediate Au concentrations in  LaAg$_{1-x}$Au$_x$Sb$_2$, for $x = 0.25, 0.75$ the growth composition of 1 : 6 : 20 (T6) was used as well. For the end compound, LaAu$_x$Sb$_2$, the data from the recent Ref. \cite{xia20a} are used when appropriate.

Cu-K$_{\alpha}$ x-ray diffraction patterns were taken using a Rigaku Miniflex-II diffractometer. The crystals were ground and the powder was mounted on a low-background single-crystal silicon plate using a trace amount of Dow Corning silicone vacuum grease. The mount was spun during data collection to reduce possible effects of texture. Data taken for Rietveld refinement were collected in two overlapping blocks: $10^{\circ} \leq 2\theta \leq 48^{\circ}$ and $38^{\circ} \leq 2\theta \leq 100^{\circ}$, with the second block counted for 4-5 times longer than the first to compensate for the loss of scattered intensity at higher angles due to the x-ray form factors. The two data blocks for each sample were co-refined within GSAS using a single set of structural and instrumental parameters but with independent scale factors to allow for the different counting times used. Parameters for both the primary phase and any impurity were refined. We found that the materials were easy to grind into a random powder and no texture or preferential orientation effects were observed in the residuals. The diffractometer and analysis procedures were checked using $\rm Al_2O_3$ (SRM 676a\cite{nist});  our fitted values of $a=$4.7586(2)~\AA\ and $c=$12.9903(7)~\AA\ were both 1.6(4)$\times10^{-4}$~\AA\ smaller than the values on the certificate\cite{nist}, suggesting a small but statistically significant mis-calibration of the instrument. The fitted lattice parameters given in the analysis that follows do not include this correction.

Chemical analysis of the crystals was performed using an Oxford Instruments energy-dispersive x-ray spectroscopy (EDS) system on a Thermo Scientific Teneo scanning electron microscope. The measurements were performed on polished $ab$ surfaces of single crystals with four to eight points taken for every sample.

Standard, linear 4-probe ac resistivity was measured on bar - shaped samples of LaAg$_{1-x}$Au$_x$Sb$_2$ in two arrangements: $I || ab$ and, when needed, $I || c$. The size of the samples was 1.5 - 2 mm length, 0.2 - 0.4 mm width and about 0.1 mm thickness. The frequency used was 17 Hz, typical current values were 3 mA for in-plane electrical transport and 5 mA for the $c$ - axis measurements.  The contact resistances between the leads and the samples were below 1Ω. Based on our experience with the LaAu$_x$Sb$_2$ samples with similar size and contact resistance, \cite{xia20a} we do not expect the heating effects to be observed either at ambient pressure or in the	pressure cell environment. The measurements were performed using the ACT option of a  Quantum Design Physical Property Measurement System (PPMS).

For selected samples, resistivity measurements under pressure were performed in a hybrid, BeCu / NiCrAl piston - cylinder pressure cell (modified version of the one used in Ref. \cite{bud86a}) in the temperature environment provided by a PPMS instrument. A 40 : 60 mixture of light mineral oil and n-pentane was used as a pressure-transmitting medium. This medium solidifies at room temperature in the pressure range of 30 - 40 kbar, \cite{bud86a,kim11a,tor15a} which is above the maximum pressure in this work. Elemental Pb was used as a low temperature pressure gauge.\cite{eil81a} It has been shown that in piston-cylinder pressure cells the value of pressure depends on temperature (see Ref. \cite{xia20b} for  mineral oil : n-pentane pressure medium and this particular design of the cell). Below we  use the Pb gauge pressure value. Given that the upper transition for LaAgSb$_2$, highest in the series, is at ambient pressure   at $\sim 200$ K, this may give rise to pressure differences with the values determined by Pb gauge by at most 2  kbar.

%%%%%%%%%%%%%%%%%%%%%%%%%%%%%%%%%%%%%%%%%%
\section{Results}

\subsection{Structure and substitution}

\begin{figure}
\includegraphics[width=12cm]{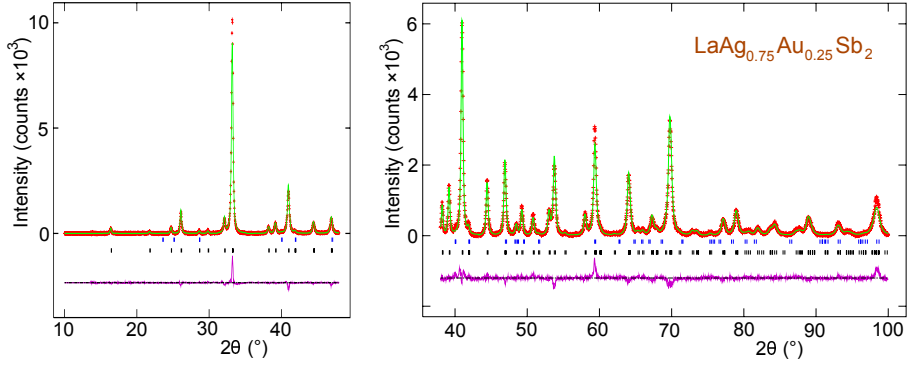}
\caption{(color online) Cu-K$_{\alpha}$ x-ray diffraction patterns for the T2 growth of  LaAg$_{0.75}$Au$_{0.25}$Sb$_2$ showing the two overlapping data blocks that were co-fitted using the GSAS/EXPGUI packages. \cite{lar00a,tob01a} The red points are the data and the green lines show the fits with the residuals shown below each fitted pattern. The Bragg markers show the positions of the reflections from (top) Sb,  and (bottom) LaAg$_{1-x}$Au$_{x}$Sb$_2$. }
\label{f1}
\end{figure}

The x-ray diffraction patterns for all  LaAg$_{1-x}$Au$_x$Sb$_2$ samples were fitted using the GSAS/EXPGUI packages.\cite{lar00a,tob01a} Small amounts of residual flux were generally observed as impurity phases and were included in the fits as necessary. Figure \ref{f1}  shows a typical x-ray diffraction data set for the T2 growth of  LaAg$_{0.75}$Au$_{0.25}$Sb$_2$ with $\sim 1$ wt.\% Sb as impuritiy. In the fit, the occupations of the La, Sb1, and Sb2 sites as well as the total occupation of the T = Ag/Au $2b$ site were fixed as 1, whereas the Ag/Au ratio was allowed to vary. As the parameter that actually contributes to the scattering from a given site in the structure is the average scattering length for that site, it was not meaningful to refine both the Au/Ag ratio and a possible vacancy level using a single measurement (our x-ray diffraction patterns) of the average scattering length. The same (reduced) average scattering could be constructed from Au-only + some level of vacancy, a fully occupied site with Au + some Ag, or some appropriate, and continuously variable combination of Au + Ag + vacancy. The results from Rietveld analysis of the powder x-ray data for LaAgSb$_2$ and five LaAg$_{1-x}$Au$_x$Sb$_2$ samples are listed in Tables \ref{T1} and \ref{T2} in the Appendix. The EDS results for LaAgSb$_2$ and four LaAg$_{1-x}$Au$_x$Sb$_2$ samples are presented in Table \ref{T3} in the Appendix. The values in the table are the average of the measurements taken at between four and eight different places on the samples' surfaces, standard deviations are listed in the parentheses. 

\begin{figure}
\includegraphics[width=10cm]{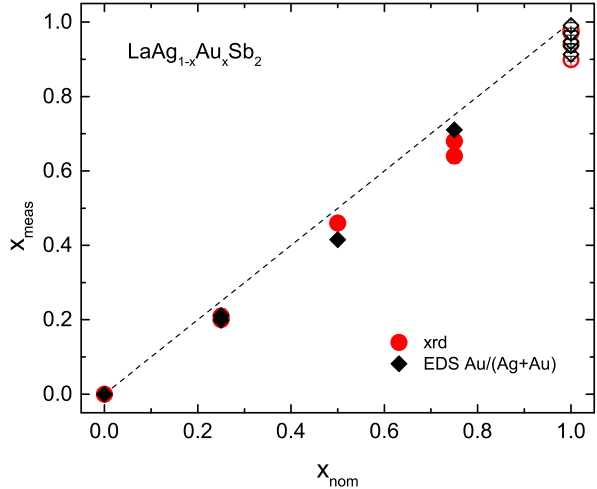}
\caption{(color online) Measured vs nominal values of Au concentration $x$ in  LaAg$_{1-x}$Au$_x$Sb$_2$ (filled symbols). Double filled red circles for $x_{nom} = 0.25$ (not clearly discerned on the plot) and 0.75, as well as double filled black rhombi for  $x_{nom} = 0.25$ correspond to T2 and T6 growths, see Materials and Methods section. These data are presented in the Appendix in a tabular form (tables \ref{T2}, \ref{T3}). For example of two red circles at $x_{nom} = 0.75$, higher corresponds to T2 and lower to T6 growth. Data for  LaAu$_x$Sb$_2$ \cite{xia20a} (open symbols) are added for the reference, here again multiple symbols correspond to different initial La : Au : Sb growth compositions. \cite{xia20a} Dashed line corresponds to $x_{meas} = x_{nom}$.}
\label{f2}
\end{figure}

Analysis of the x-ray diffraction as well as EDS results show (Fig. \ref{f2}) that the measured Ag/Au ratio  deviates from the nominal with the experimental points for $x_{meas}$  being slightly below the $x_{meas} = x_{nom}$ line with $x_{meas}/x_{nom} = 0.88 \pm 0.02$ and $0.90 \pm 0.03$ for x-ray diffraction and EDS results respectively. In the rest of the text we will use $x$ - values determined from the x-ray diffraction.

\begin{figure}
\includegraphics[width=10cm]{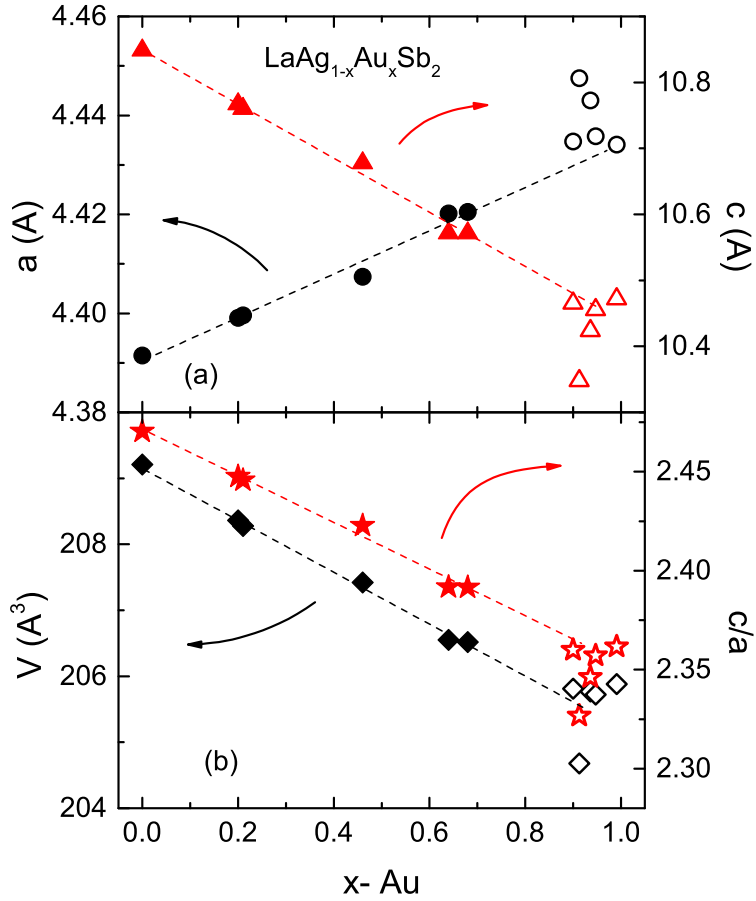}
\caption{(color online) Lattice parameters, unit cell volume and $c/a$ ratio vs  Au concentration in  LaAg$_{1-x}$Au$_x$Sb$_2$ determined from Rietveld refinement (filled symbols).  Data for  LaAu$_x$Sb$_2$ \cite{xia20a} (open symbols) are added for the reference. Dashed lines are guide to the eye. }
\label{f3}
\end{figure}

The lattice parameters, unit cell volume and the $c/a$ ratio as a function of Au substitution are presented in Fig. \ref{f3}. All these quantities have an approximately linear dependence of $x$, in good agreement with Ref. \cite{mas14a}. 

\subsection{CDW at ambient pressure}

\begin{figure}
\includegraphics[width=7cm]{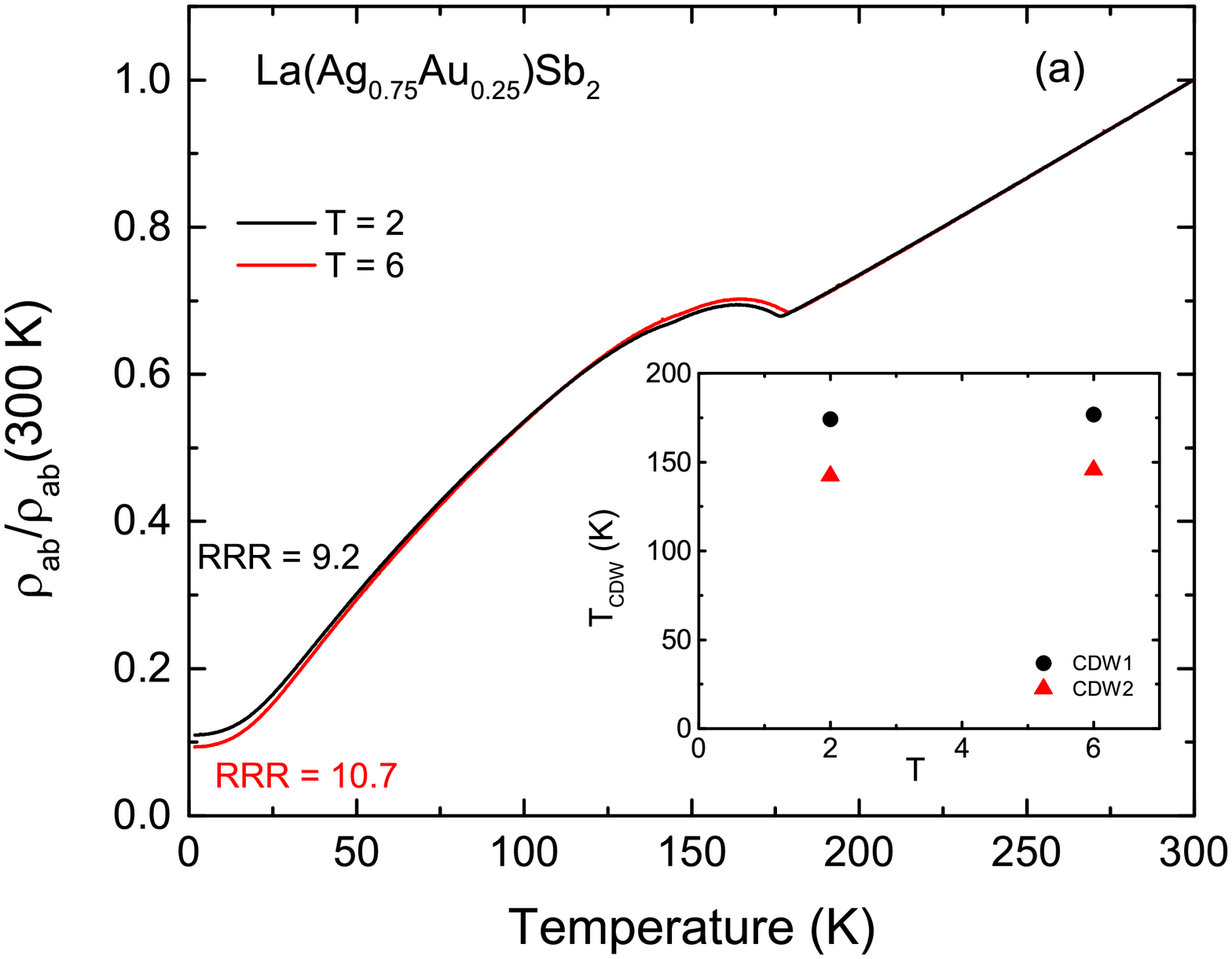}
\includegraphics[width=7cm]{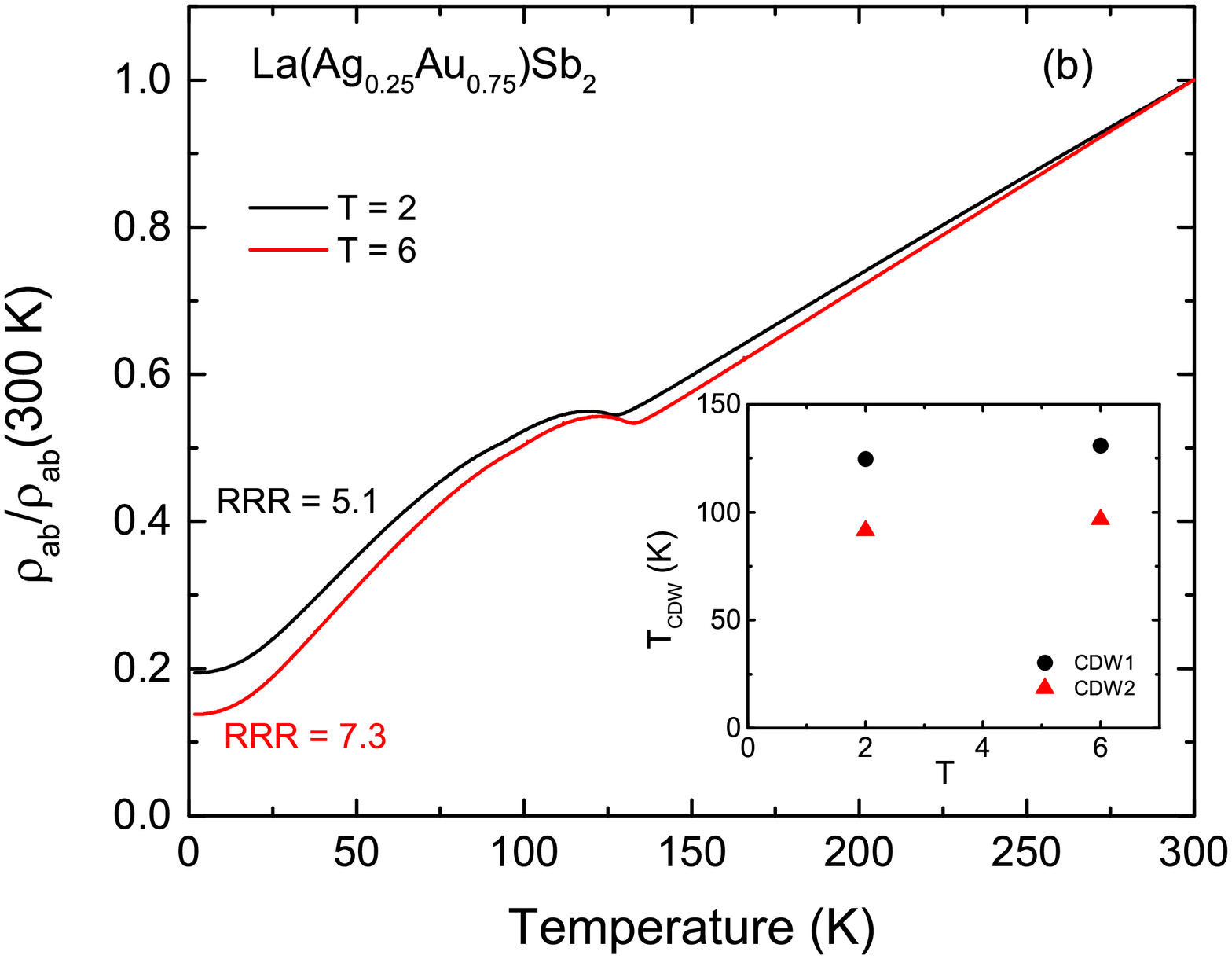}
\caption{(color online) Normalized in-plane resistivity, $\rho_{ab}/\rho_{ab}(300~\text{K})$ and the ordering temperatures CDW1 and CDW2 for T2 and T6 samples with nominal compositions   (a) LaAg$_{0.75}$Au$_{0.25}$Sb$_2$ and (b)  LaAg$_{0.25}$Au$_{0.75}$Sb$_2$. }
\label{resAB}
\end{figure}

Whereas in the case of LaAu$_x$Sb$_2$ the CDW temperatures were significantly affected by initial growth compositions, \cite{xia20a} this appears to be not so critical for LaAg$_{1-x}$Au$_x$Sb$_2$ with the intermediate Au compositions. For the nominal  LaAg$_{0.75}$Au$_{0.25}$Sb$_2$ and  LaAg$_{0.25}$Au$_{0.75}$Sb$_2$ samples the difference between T2 and T6 initial compositions in the XRD-refined Au concentrations  is 0.01 - 0.04 (5 - 6\%) (Table \ref{T2}) and in the CDW ordering temperatures 3 - 6 K (2 - 5\%), with the difference, not surprisingly,  being larger for the latter samples with higher Au concentration. The normalized in-plane resistivity and the CDW ordering temperatures for T2 and T6 samples of  LaAg$_{0.75}$Au$_{0.25}$Sb$_2$ and  LaAg$_{0.25}$Au$_{0.75}$Sb$_2$ are shown in Fig. \ref{resAB}(a) and (b) respectively.

\begin{figure}
\includegraphics[width=10cm]{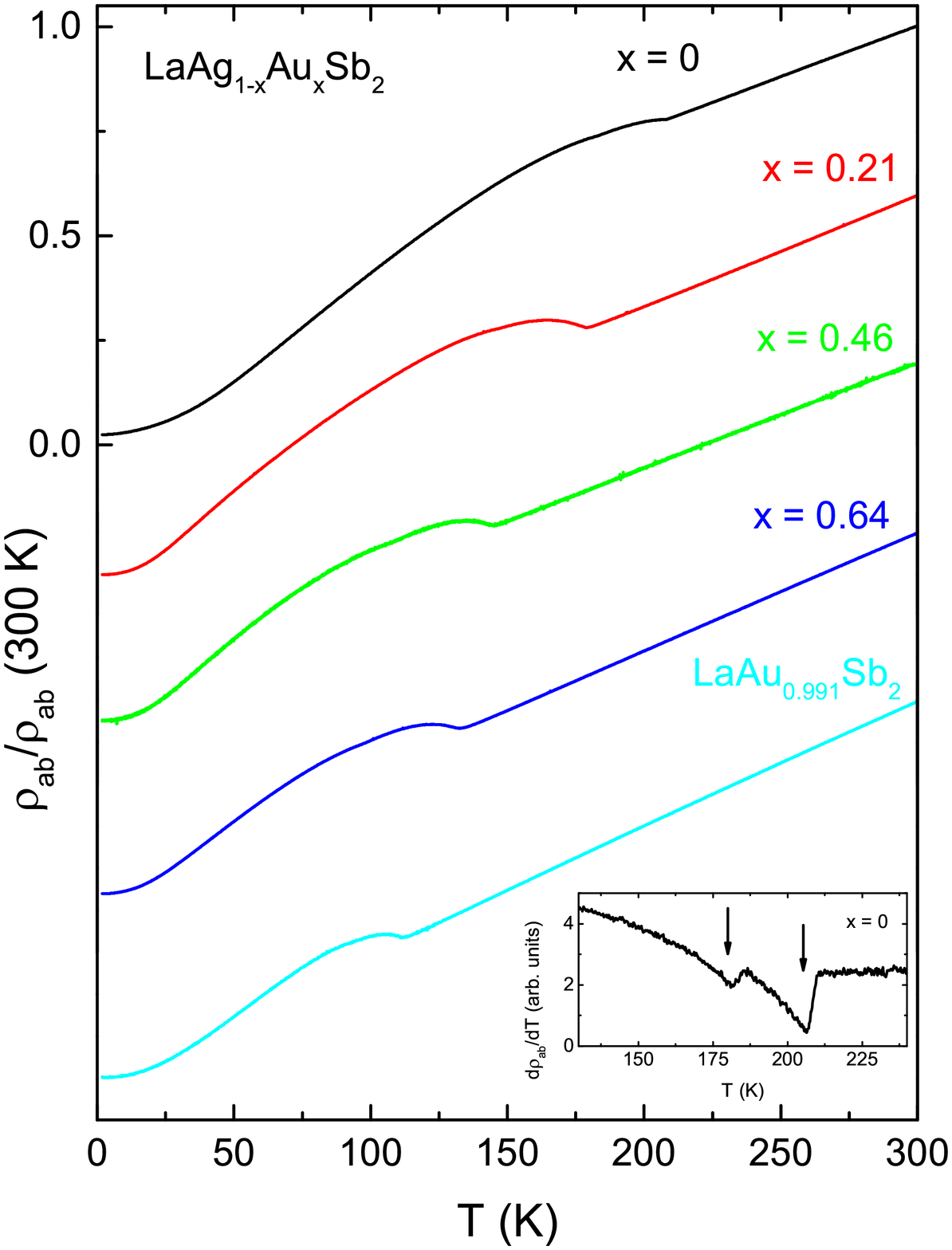}
\caption{(color online) Normalized in-plane resistivity for  LaAg$_{1-x}$Au$_{x}$Sb$_2$. Data are vertically shifted  for clarity.  The curve for  LaAu$_{0.991}$Sb$_2$ \cite{xia20a} is added for the reference. The inset shows an example of $d \rho_{ab}/dT$ for LaAgSb$_2$ with two CDW transitions marked. }
\label{RT}
\end{figure}

\begin{figure}
\includegraphics[width=10cm]{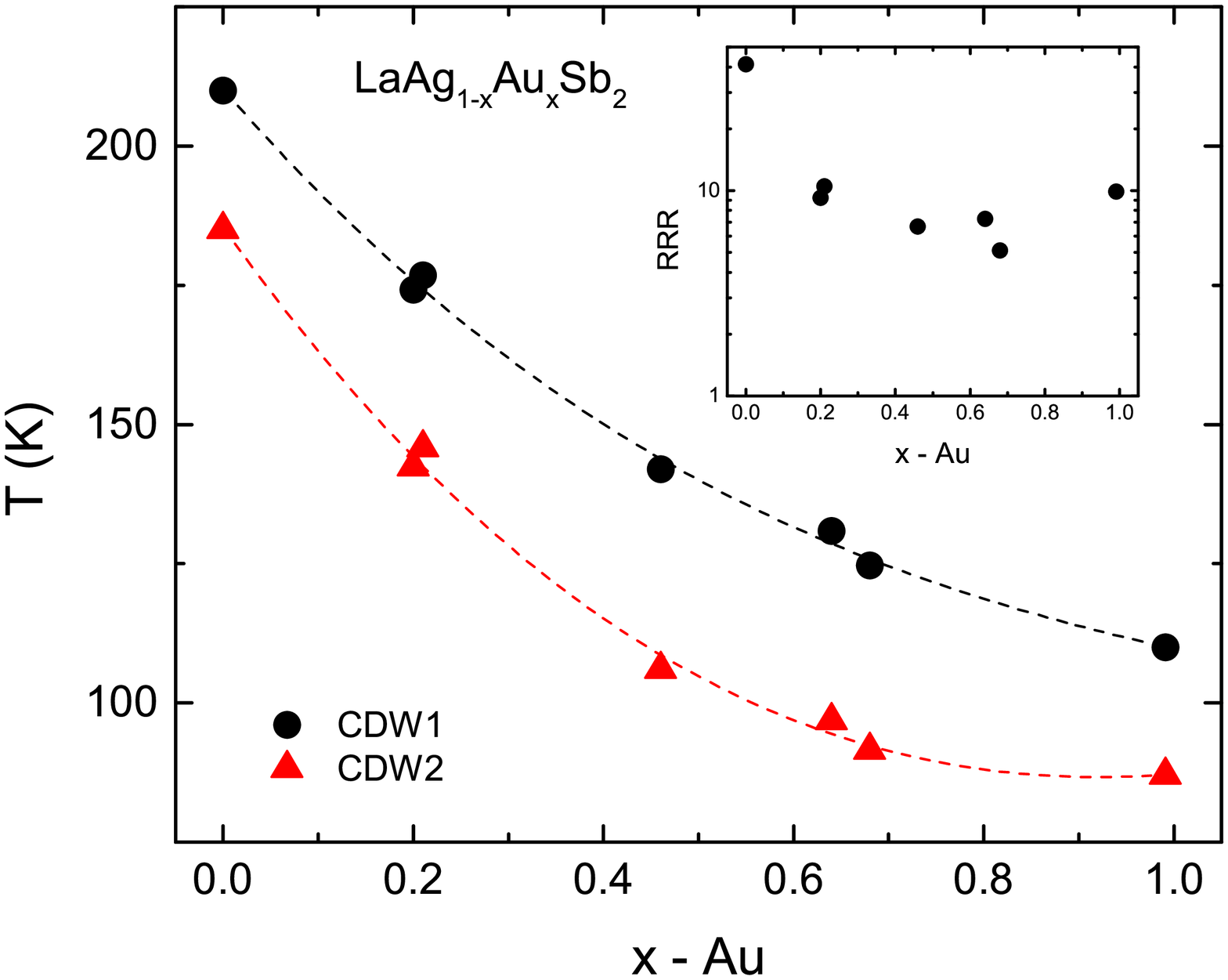}
\caption{(color online) Transition temperatures, CDW1 and CDW2, as a function of $x$ in  LaAg$_{1-x}$Au$_{x}$Sb$_2$. Data are vertically shifted  for clarity. Dashed lines are guide for the eye. The inset shows residual resistivity ratio, $RRR = \rho_{ab} (300~\text{K})/\rho_{ab}(1.8~\text{K})$ as a function of $x$ - Au. }
\label{xT}
\end{figure}

The overall evolution of the in-plane resistivity of   LaAg$_{1-x}$Au$_{x}$Sb$_2$ is shown in Fig. \ref{RT}. CDW transition temperatures decrease with Au substitution. The suppression of $T_{CDW1}$ is in fair agreement with the prior results. \cite{mas14a} The ambient pressure $x - T$ phase diagram based on the data of Figs \ref{resAB} and \ref{RT} is presented in Fig \ref{xT}.  The $T_{CDW}$ values were determined from extrema in the $d\rho_{ab}/dT$ data; an example of which is shown in the inset to figure \ref{RT}.

On going from LaAgSb$_2$ to LaAuSb$_2$, based on electrical transport data, both CDW transitions persist, but both of them are suppressed by $\sim 100$ K without a significant change of the value of $T_{CDW1} - T_{CDW2}$. The $T_{CDW1}(x)$ and $T_{CDW2}(x)$ behavior has an upward curvature. Most likely the disorder induced by substitution contributes to additional suppression of CDW transition temperatures, although to the extent significantly smaller than e.g. in 2H - TaSe$_{2-x}$S$_x$. \cite{lil17a} The presence of substitutional disorder is, expectedly, seen in the evolution of residual resistivity ratio ($RRR = \rho_{ab} (300~\text{K})/\rho_{ab}(1.8~\text{K})$) with Au substitution (Fig. \ref{xT}, inset), which shows a broad local minimum for intermediate substitution values.  Similar moderate but visible effect of substitutional disorder was observed in studies of superconducting transition temperature in Y$_x$Lu$_{1-x}$Ni$_2$B$_2$C. \cite{fuc02a} A clearer example can be found in a similar, isoelectronic substitution in the Mn(Pt$_{1-x}$Pd$+x$)$_5$P series. \cite{sla22a}

\subsection{CDW under pressure}

\begin{figure}
\includegraphics[width=10cm]{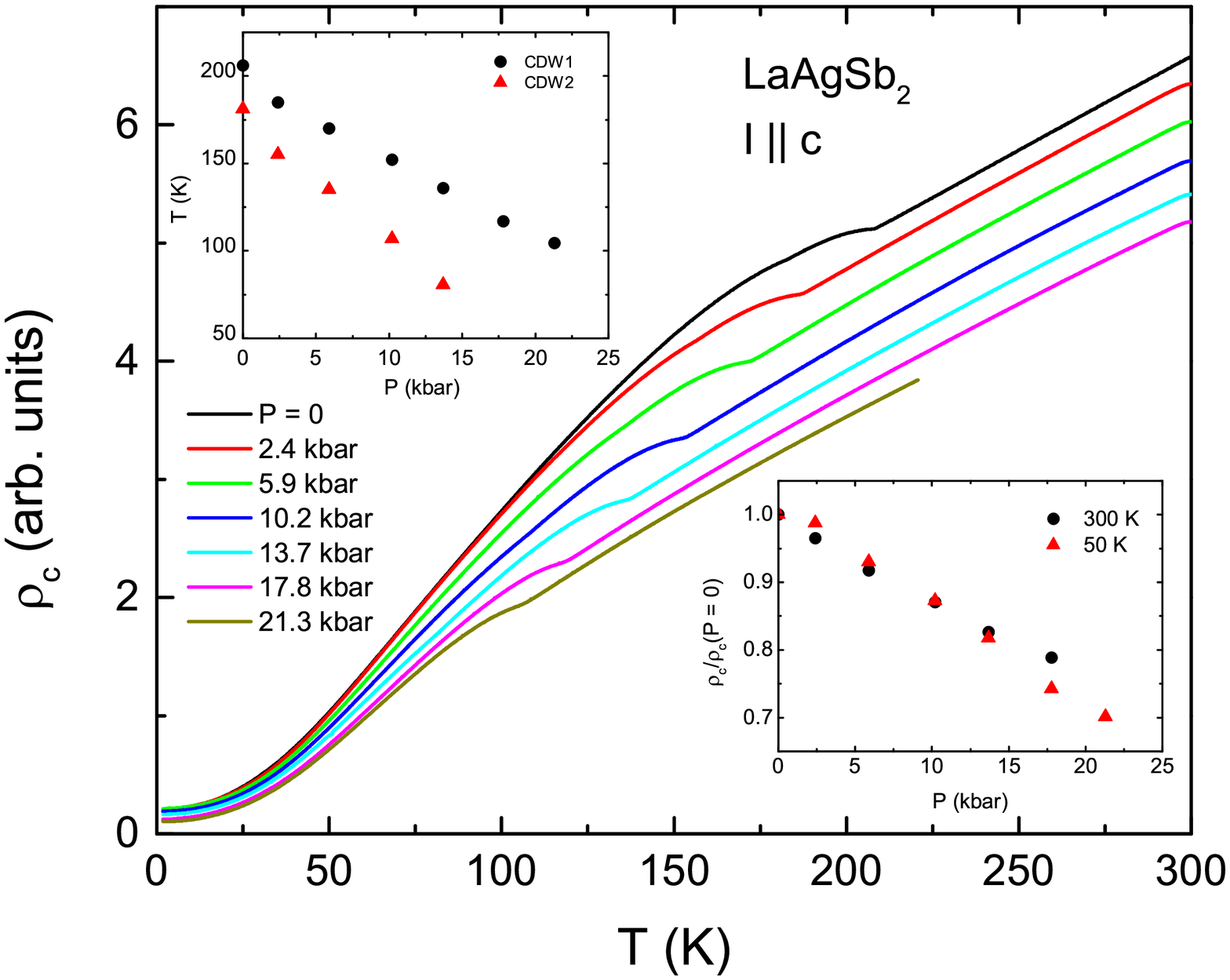}
\caption{(color online) Temperature-dependent $c$-axis resistivity of LaAgSb$_2$ measured at different applied pressures. Upper inset: CDW transition temperatures, $T_{CDW1}$ and $T_{CDW2}$, as a function of pressure. Lower inset: normalized resistivity at 300 K and 50 K as a function of pressure. }
\label{Px0}
\end{figure}

Note that often (see above) both CDW1 and CDW2 are reasonably well discerned in the derivatives of in-plane resistivity, with the feature associated with CDW2 being less pronounced.  Having in mind that (at least in LaAgSb$_2$ \cite{son03a}) the CDW2 wavevector is along the $c$-axis, the $\rho_c$ measurements provide better identification of the $T_{CDW2}$ with $T_{CDW1}$ still being strong. Therefore the measurements for LaAgSb$_2$ under pressure were performed in $I \| c$, $H \| ab$ geometry.

Main panel of the Fig. \ref{Px0} presents $c$ - axis resistivity data taken for LaAgSb$_2$ at different pressures. The overall resistivity is suppressed as pressure increases. The CDW transitions are moving down in temperature.  The insets help to quantify above statement. The relative change of the c-axis resisitivity under pressure, is similar for both temperatures presented, 300 K and 50 K, $1/\rho_c(0) \cdot d \rho_c/dP = - (0.012 -0.014)$~kbar$^{-1}$.  Both CDW temperatures decrease under in a close to linear fashion, with the derivatives $dT_{CDW1}/dP = - 4.6 \pm 0.2$~K/kbar and $dT_{CDW2}/dP = - 7.0 \pm 0.3$~K/kbar. Simple, linear, extrapolation suggests that CDW2 will be suppressed to 0 K at $\sim 25$~kbar and CDW1 at $\sim 43$~kbar. The observed $T_{CDW}$ derivatives are consistent with the published values of $- (4.3 - 5.1)$ K/kbar for CDW1 \cite{bud06a,tor07a,aki21a} and - 8.0~K/kbar for CDW2. \cite{aki21a}

\begin{figure}
\includegraphics[width=10cm]{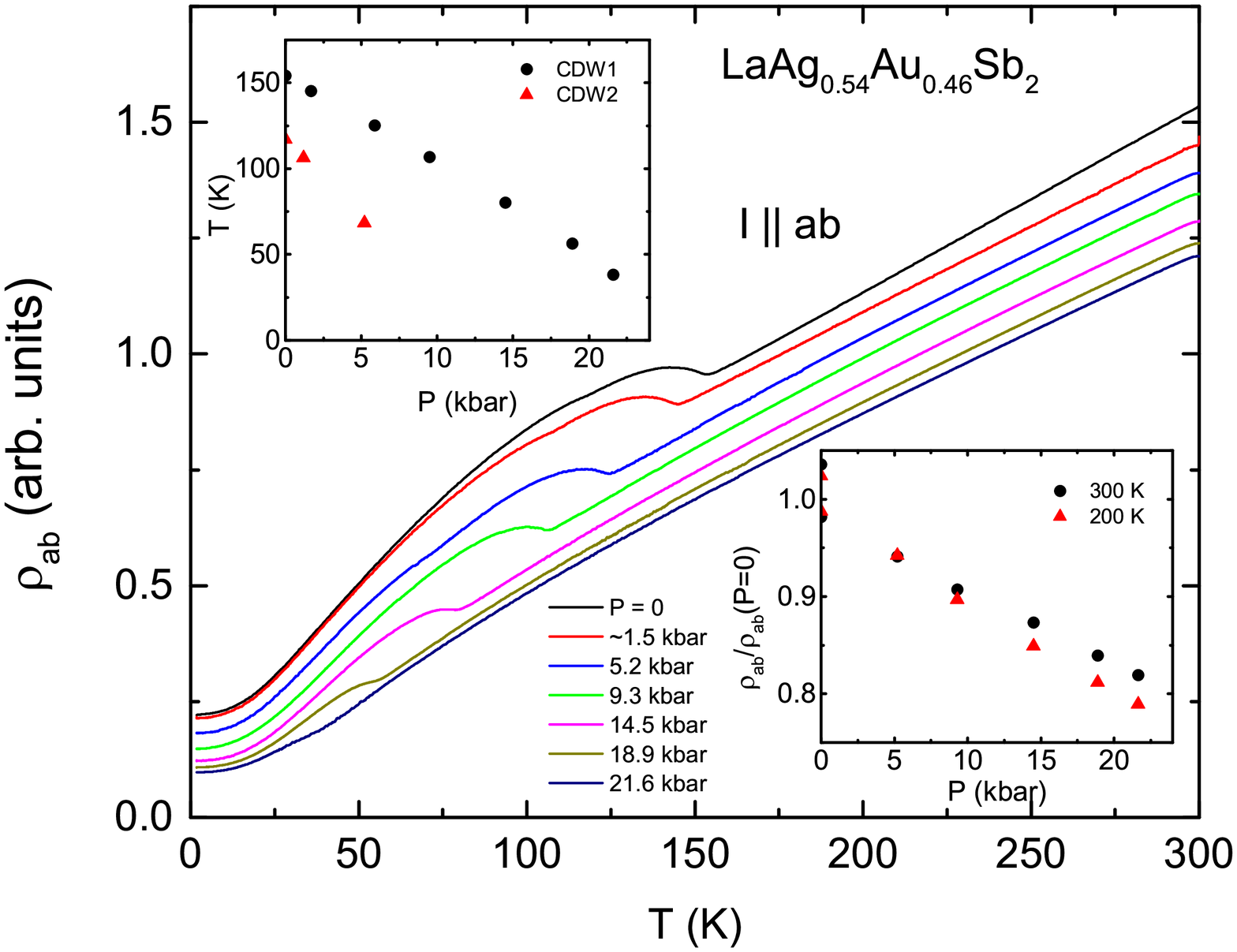}
\caption{(color online) Temperature-dependent in-plane resistivity of LaAg$_{0.54}$Au$_{0.46}$Sb$_2$ measured at different applied pressures. Upper inset: CDW transition temperatures, $T_{CDW1}$ and $T_{CDW2}$, as a function of pressure. Lower inset: normalized resistivity at 300 K and 200 K as a function of pressure. Note, some minor corrections of pressure values, following Ref. \cite{xia20a} were applied. }
\label{Px50}
\end{figure}

\begin{figure}
\includegraphics[width=10cm]{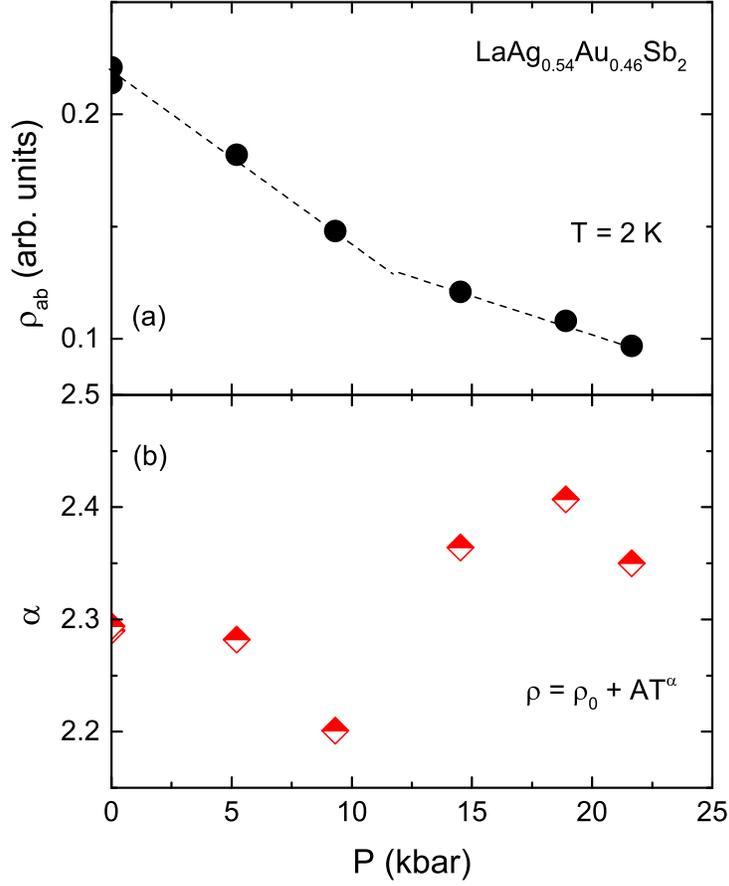}
\caption{(color online) Pressure dependence of (a) in-plane resistivity at 2 K (dashed lines are guide to the eye); (b) exponent $\alpha$ in $\rho = \rho_0 + AT^{\alpha}$ fit of low temperature  resistivity (fit performed between 1.8 K and 20 K) for LaAg$_{0.54}$Au$_{0.46}$Sb$_2$ .}
\label{PLT}
\end{figure}

In order to extend the pressure dependence of $T_{CDW1}$ and $T_{CDW2}$ across the substitutional series, a similar data set (but for $\rho_{ab}$) for LaAg$_{0.54}$Au$_{0.46}$Sb$_2$ (nominal LaAg$_{0.5}$Au$_{0.5}$Sb$_2$) is presented in Fig. \ref{Px50}. The relative change in the in-plane resistivity at 300 K and 200 K is $1/\rho_{ab}(0) \cdot d \rho_{ab}/dP = - (0.009 -0.01)$~kbar$^{-1}$, the same as that in LaAgSb$_2$. \cite{tor07a} The initial pressure derivatives of the CDW transitions are $dT_{CDW1}/dP = - 4.9 \pm 0.1$~K/kbar and $dT_{CDW2}/dP = - 9.4 \pm 0.1$~K/kbar. For CDW2 simple, linear, extrapolation yields $\sim 12.5$~kbar as a critical pressure of complete suppression of CDW2. Since we cannot observe any distinguishable feature in $d\rho_{ab}/dT$ data at 9.3 kbar below 50 K (see Fig. \ref{Ader} in the Appendix),  it is possible that the $T_{CDW2}(P)$ behavior is super-linear  and the critical pressure for CDW2 is lower than $\sim 12.5$~kbar obtained from the linear extrapolation. Alternatively, the feature associated with CDW2 could be suppressed so much, that it cannot be detectable within our signal-to-noise ratio and digital differentiation. The data in Fig. \ref{PLT} potentially favors the former possibility. $T_{CDW1}(P)$ dependence has some curvature, the data extrapolate to the value of the critical pressure of $\sim 26$~kbar.

Since for LaAg$_{0.54}$Au$_{0.46}$Sb$_2$ CDW2 appears to be suppressed to 0~K within our pressure range, we examine if this suppression has any bearing on the low temperature electrical transport. Indeed, the zero applied field data in figures  \ref{PLT}(a) and  \ref{PLT}(b) show changes in behavior near 9.3 kbar with the power law exponent, $\alpha$, having the clearest signature of a possible transition near the 9.3 kbar.  So most probably the value of the critical pressure for CDW2 is around 9.3 kbar. The changes observed are rather subtle, however the features associated with CDW suppression in LaAgSb$_2$ \cite{aki21a} and LaAu$_x$Sb$_2$ \cite{xia20a} were subtle as well.

%%%%%%%%%%%%%%%%%%%%%%%%%%%%%%%%%%%%%%%%%%
\section{Discussion}

\begin{figure}
\includegraphics[width=7cm]{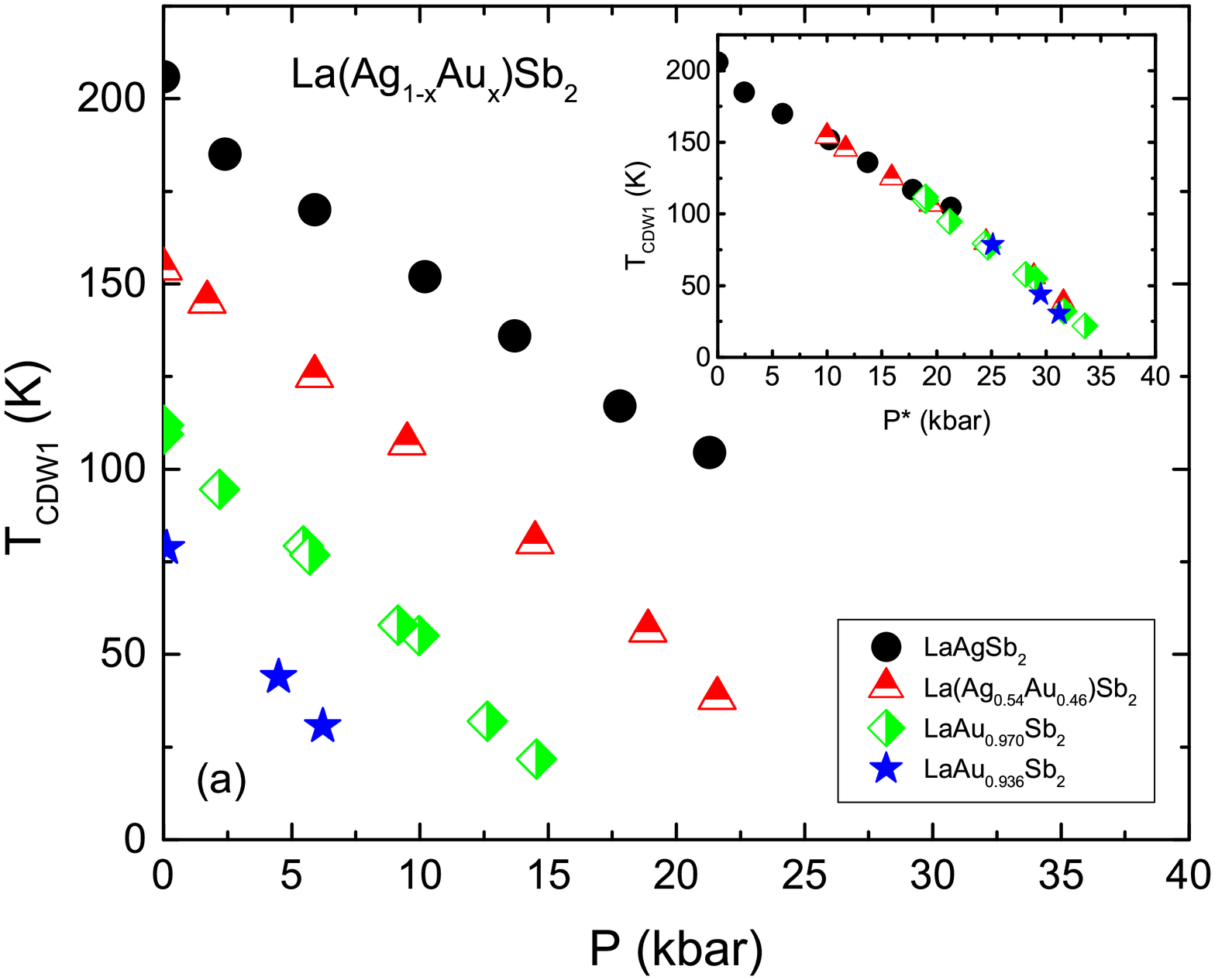}
\includegraphics[width=7cm]{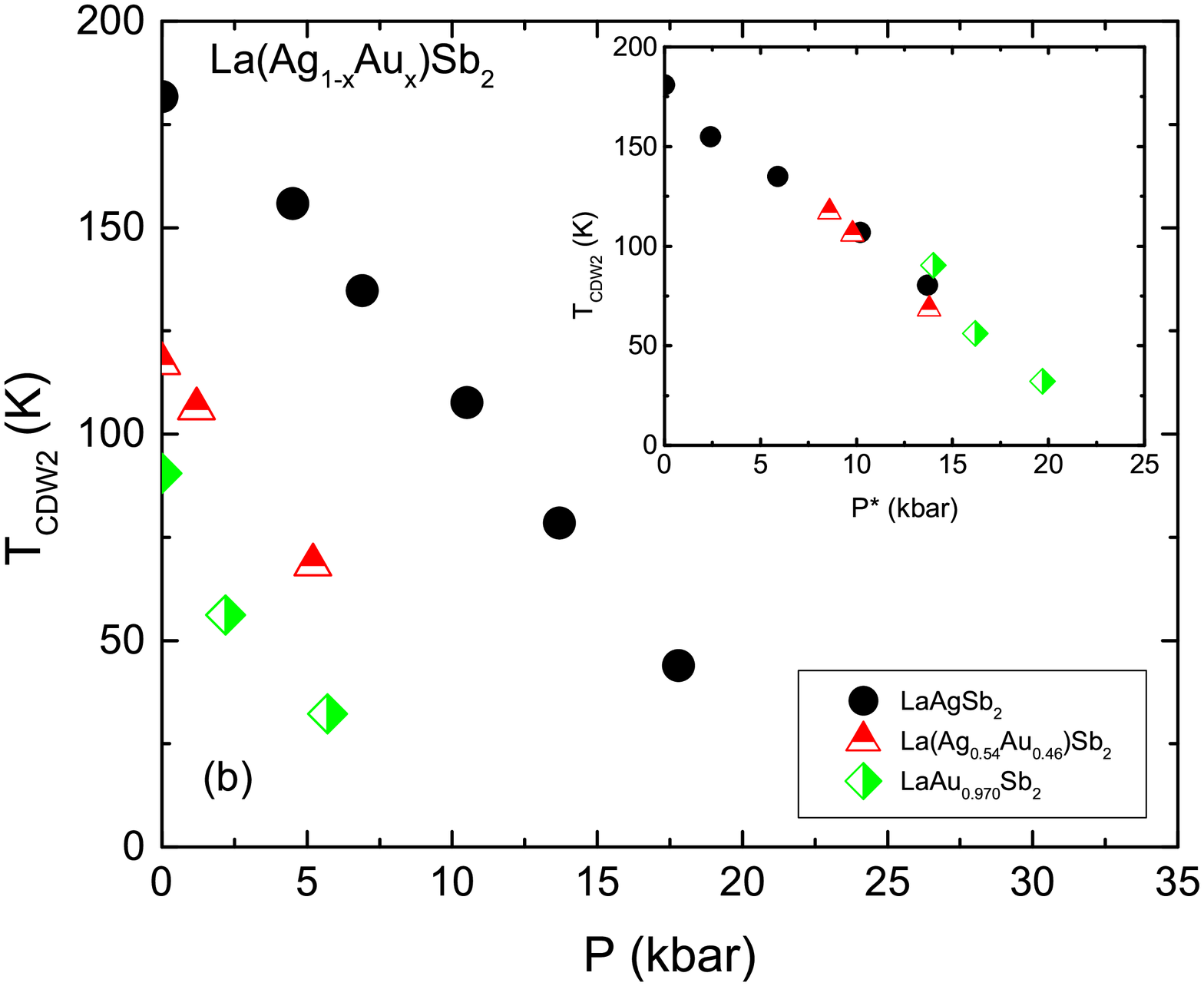}
\caption{(color online) Pressure dependence of (a) $T_{CDW1}$ and (b) $T_{CDW2}$ for different La(Ag$_{1-x}$Au$_x$)Sb$_2$. shown on the same plot. Data for LaAu$_x$Sb$_2$ are taken from Ref. \cite{xia20a}. Insets: the same data positioned on universal lines by horizontal shifts, $\Delta P$.}
\label{T12PM}
\end{figure}

The pressure dependence of the CDW transitions of LaAgSb$_2$, La(Ag$_{0.54}$Au$_{0.46}$)Sb$_2$ and LaAu$_x$Sb$_2$ samples is shown in Fig. \ref{T12PM}. For different members of the family the behavior is very similar. It is noteworthy that the CDW2 suppression rates are almost a  factor of 2 higher than those for CDW1. This is possibly due to different effect of pressure on the nesting features along $a$- and $c$-axes (note that for LaAgSb$_2$ the CDW wave-vectors were found to be (0.026~0~0) and (0~0~0.16) for CDW1 and CDW2 respectively \cite{son03a}).

The $T_{CDW}(P)$ data could be combined on the same universal line by horizontal shift of the data (as shown in the insets to Fig. \ref{T12PM}). This universal behavior suggest equivalence of the chemical and physical pressure. The approximate scaling is 2 kbar~$\sim 0.1~x$-Au for CDW1 and slightly smaller pressure shift per 0.1~$x$-Au for CDW2. We recall that in a similar way the $P - T$ phase diagrams for Ba(Fe$_{1-x}$Ru$_x$)$_2$As$_2$ with different value of $x$ were combined to form a universal phase diagram by $\Delta P$ shifts with 30 kbar $\sim 0.1~x$-Ru. \cite{kim11a} In contrast, the pressure and substitution data in the La$_{1-x}$$R_x$AgSb$_2$ ($R$ = Ce, Nd) series \cite{tor07a} cannot be combined on the same line by $\Delta P$ shifts. Apparently rare earth and transition metal substitutions in LaAgSb$_2$ affect the pressure derivatives of CDW transition temperatures in different manner, with $R$-substituted compounds having higher (and $R$-dependent) suppression rates. Of course, whereas both Ag/Au and La/R substitutions are isoelectronic, substitution of Ce or Nd for La brings local moment magnetism that subsitution of Au for Ag does not.

\begin{figure}
\includegraphics[width=12cm]{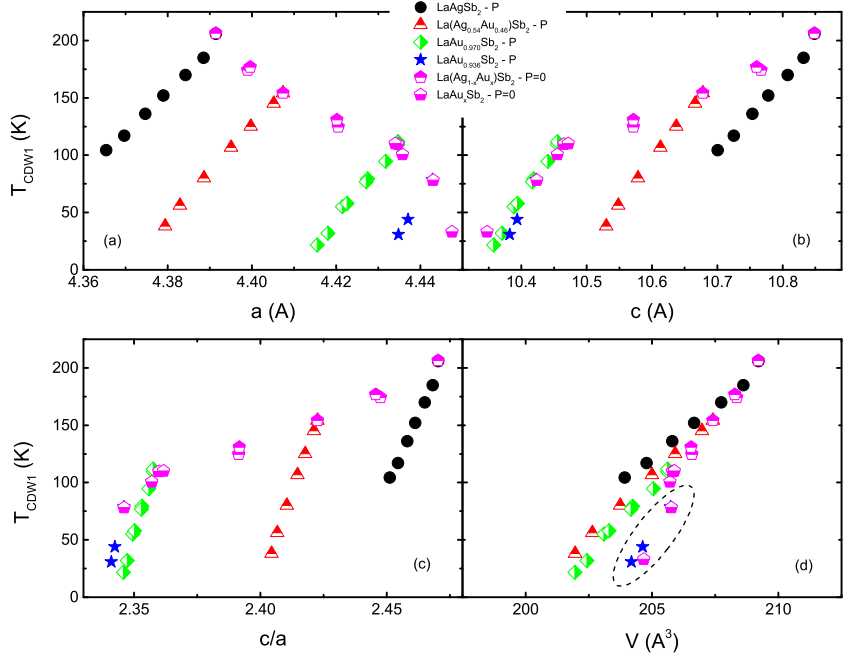}
\caption{(color online) CDW1 transition temperature for La(Ag$_{1-x}$Au$_x$)Sb$_2$ and LaAu$_x$Sb$_2$ \cite{xia20a} at ambient and high pressure as a function of (a) $a$; (b) $c$; (c) $c/a$; and (d) $V$ structural parameters. Encircled points discussed in the text.}
\label{stru1}
\end{figure}

To gain some further insight on which structural parameter could be of importance for change of the CDW temperature under pressure and with Au substitution we plot $T_{CDW1}$ as a function of basic structural parameters, $a, c, c/a$ and $V$ in Fig. \ref{stru1}. For ambient pressure data the structural parameters obtained from the Rietveld refinement are used. For the high pressure data the structural parameters were obtained from the $P = 0$ values using LaAgSb$_2$ elastic constants from Ref. \cite{bud06a} and assuming that their change within the La(Ag$_{1-x}$Au$_x$)Sb$_2$ series is insignificant.

\begin{figure}
\includegraphics[width=10cm]{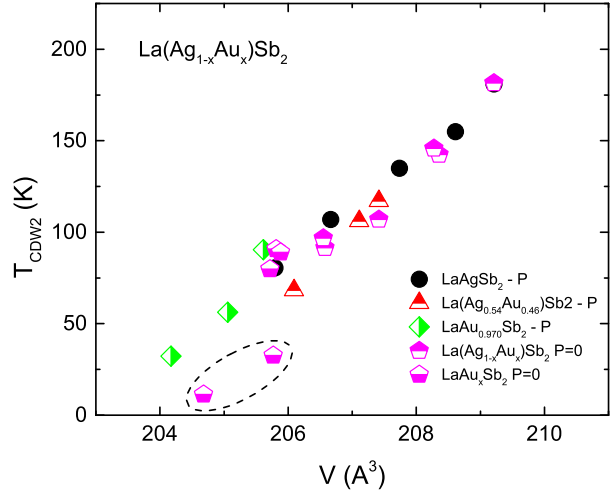}
\caption{(color online) CDW2 transition temperature for La(Ag$_{1-x}$Au$_x$)Sb$_2$ and LaAu$_x$Sb$_2$ \cite{xia20a} at ambient and high pressure as a function of the unit cell volume, $V$. Encircled points discussed in the text.}
\label{stru2}
\end{figure}

The data in Fig. \ref{stru1} clearly show that whereas using $a, c$, and $c/a$ as a structural parameter results in distinctly different trends for chemical and physical pressure, all the data on $T_{CDW1}$ vs $V$ plot fall fairly well on the same line. To check if this holds for CDW2 as well, we plotted $T_{CDW2}$ vs $V$ in Fig. \ref{stru2} as well. The $T_{CDW2}$ data also scale with the unit cell volume well. We have few outlier points encircled in Figs. \ref{stru1}(d) and \ref{stru2}. These point correspond to  measurably {\it off-stoichiometric}  LaAu$_x$Sb$_2$ \cite{xia20a}, whereas the rest of the data are for the compounds with stiochiometry very close to 1:1:2. That is possibly the reason for these few point being outliers. The $T_{CDW}$ vs $V$ scaling could be even better if the elastic constants measured for each compound were used, however to address this, further elastic properties measurements should be performed. It is of a surprise, that despite La(Ag$_{1-x}$Au$_x$)Sb$_2$ being tetragonal, anisotropic compounds, the chemical and physical pressure appear to be equivalent with a salient structural parameter being the unit cell volume (that lacks any anisotropic information). Hopefully further band structure calculations will be able to address this issue.

%%%%%%%%%%%%%%%%%%%%%%%%%%%%%%%%%%%%%%%%%%
\section{Summary}

Study of compounds of the La(Ag$_{1-x}$Au$_x$)Sb$_2$ family at ambient and high pressure show that both CDW transitions are suppressed with Au substitution and under pressure in a manner that indicate equivalence of chemical and physical pressure in this series with the unit cell volume being a suitable structural control parameter and with suppression rates being different for CDW1 and CDW2. Such equivalence of physical and chemical pressure  was not observed in the La$_{1-x}$$R_x$AgSb$_2$ series. \cite{tor07a}. Different CDW suppression rates probably reflect the fact that (at least for LaAgSb$_2$) the CDW wave-vectors are orthogonal, along $a$- and $c$-axis for CDW1 and CDW2 respectively.

Additionally, for La(Ag$_{0.54}$Au$_{0.46}$)Sb$_2$ anomalies in low temperature electrical transport  were observed in the pressure range where CDW2 is completely suppressed.

%%%%%%%%%%%%%%%%%%%%%%%%%%%%%%%%%%%%%%%%%%

%%%%%%%%%%%%%%%%%%%%%%%%%%%%%%%%%%%%%%%%%%
\begin{acknowledgments}

We thank Warren E. Straszheim for help with the EDS measurements, and Elena Gati and Raquel A. Ribeiro for useful discussions. PCC would like to acknowledge Paul Delvaux for ongoing inspiration. Work at the Ames National Laboratory was supported by the U.S. Department of Energy, Office of Science, Basic Energy Sciences, Materials Sciences and Engineering Division. The Ames National Laboratory is operated for the U.S. Department of Energy by Iowa State University under contract No. DE-AC02-07CH11358. L.X. was supported, in part, by the W. M. Keck Foundation. Much of this work was carried out while D.H.R. was on sabbatical at Iowa State University and Ames Laboratory and their generous support (again under under contract No. DE-AC02-07CH11358) during this visit is gratefully acknowledged. D.H.R. was supported as well by Fonds Qu\'eb\'ecois de la Recherche sur la Nature et les Technologies.

\end{acknowledgments}

%%%%%%%%%%%%%%%%%%%%%%%%%%%%%%%%%%%%%%%%%%

\clearpage

\appendix
\section{Rietveld refinement and EDS results}

This Appendix contains tables with the results of Rietveld refinements and EDS chemical analysis of the LaAg$_{1-x}$Au$_x$Sb$_2$ samples. Data for LaAu$_x$Sb$_2$\cite{xia20a} in the Table \ref{T1}  are added for comparison.

\begin{table*}[h]

\caption{\label{T1}Lattice parameters of LaAg$_{1-x}$Au$_x$Sb$_2$ samples (labels in parentheses indicate initial growth compositions, see Experimental details section for more details).\\}
\begin{tabular}{cccc}
\hline \hline
Sample&$a$ (\AA)&$c$ (\AA)&$V$ (\AA$^3$)\\ \hline
LaAgSb$_2$ (T2)&4.3915(1)&10.8485(4)&209.21(1)\\
LaAg$_{0.75}$Au$_{0.25}$Sb$_2$ (T2)&4.3991(1)&10.7669(4)&208.36(1)\\
LaAg$_{0.75}$Au$_{0.25}$Sb$_2$ (T6)&4.3996(1)&10.7601(4)&208.28(1)\\
LaAg$_{0.5}$Au$_{0.5}$Sb$_2$ (T2)&4.4074(2)&10.6777(6)&207.42(2)\\
LaAg$_{0.25}$Au$_{0.75}$Sb$_2$ (T2)&4.4205(2)&10.5715(5)&206.52(2)\\
LaAg$_{0.25}$Au$_{0.75}$Sb$_2$ (T6)&4.4202(1)&10.5716(4)&206.55(1)\\
LaAu$_x$Sb$_2$ (T2)\cite{xia20a}&4.4430(2)&10.4237(4)&205.77(1)\\
LaAu$_x$Sb$_2$ (T6)\cite{xia20a}&4.4347(1)&10.4653(3)&205.88(1)\\
\hline \hline
\end{tabular}

\end{table*}

\begingroup
\begin{table*}[h]
\caption{\label{T2}Atomic coordinates, occupancy, and isotropic displacement parameters of LaAg$_{1-x}$Au$_x$Sb$_2$ samples (labels in parentheses indicate initial growth compositions, see Experimental details section for more details)\\}
\renewcommand{\arraystretch}{0.75}
\begin{tabular}{cccccccc}
\hline \hline
Sample&atom&site&x&y&z&occupancy&$U_{eq}$\\ \hline 

LaAgSb$_2$ (T2)&La&2c&0.25&0.25&0.2397(1)&1&0.0260(5)\\
&Ag&2b&0.75&0.25&0.5&1&0.0297(6)\\
&Sb1&2a&0.75&0.25&0&1&0.0275(5)\\
&Sb2&2c&0.25&0.25&0.6691(2)&1&0.0275(5)\\ \hline

LaAg$_{0.75}$Au$_{0.25}$Sb$_2$ (T2)&La&2c&0.25&0.25&0.2424(2)&1&0.0335(6)\\
&Ag&2b&0.75&0.25&0.5&0.80(1)&0.033(1)\\
&Au&2b&0.75&0.25&0.5&0.20(1)&0.033(1)\\
&Sb1&2a&0.75&0.25&0&1&0.0318(7)\\
&Sb2&2c&0.25&0.25&0.6696(2)&1&0.0318(7)\\ \hline

LaAg$_{0.75}$Au$_{0.25}$Sb$_2$ (T6)&La&2c&0.25&0.25&0.2418(2)&1&0.0281(7)\\
&Ag&2b&0.75&0.25&0.5&0.79(2)&0.027(1)\\
&Au&2b&0.75&0.25&0.5&0.21(2)&0.027(1)\\
&Sb1&2a&0.75&0.25&0&1&0.0275(7)\\
&Sb2&2c&0.25&0.25&0.6700(2)&1&0.0275(7)\\ \hline

LaAg$_{0.5}$Au$_{0.5}$Sb$_2$ (T2)&La&2c&0.25&0.25&0.2448(2)&1&0.0366(7)\\
&Ag&2b&0.75&0.25&0.5&0.54(2)&0.042(1)\\
&Au&2b&0.75&0.25&0.5&0.46(1)&0.042(1)\\
&Sb1&2a&0.75&0.25&0&1&0.0357(8)\\
&Sb2&2c&0.25&0.25&0.6700(2)&1&0.0357(8)\\ \hline

LaAg$_{0.25}$Au$_{0.75}$Sb$_2$ (T2)&La&2c&0.25&0.25&0.2455(3)&1&0.0159(8)\\
&Ag&2b&0.75&0.25&0.5&0.32(2)&0.021(1)\\
&Au&2b&0.75&0.25&0.5&0.68(2)&0.021(1)\\
&Sb1&2a&0.75&0.25&0&1&0.0177(8)\\
&Sb2&2c&0.25&0.25&0.6999(3)&1&0.0177(8)\\ \hline

LaAg$_{0.25}$Au$_{0.75}$Sb$_2$ (T6)&La&2c&0.25&0.25&0.2453(2)&1&0.0298(6)\\
&Ag&2b&0.75&0.25&0.5&0.36(2)&0.0320(8)\\
&Au&2b&0.75&0.25&0.5&0.64(2)&0.0320(8)\\
&Sb1&2a&0.75&0.25&0&1&0.0331(6)\\
&Sb2&2c&0.25&0.25&0.6998(2)&1&0.0331(6)\\ 
\hline \hline
\end{tabular}

\end{table*}
\endgroup

\begin{table*}[h]

\caption{\label{T3}EDS results for LaAg$_{1-x}$Au$_x$Sb$_2$ samples\\}
\begin{tabular}{ccccccc}
\hline \hline
Sample&La at.\%&Ag at. \%&Au at.\%&Sb at. \%&Au/&3(Ag+Au)/\\
&&&&&(Ag+Au)&(La+Sb)\\ \hline
LaAgSb$_2$ (T2)&25.4(2)&25.9(1)&0&48.7(2)&0&1.05(2)\\
LaAg$_{0.75}$Au$_{0.25}$Sb$_2$ (T2)&25.6(1)&20.4(2)&5.08(7)&48.9(1)&0.199(5)&1.03(6)\\
LaAg$_{0.75}$Au$_{0.25}$Sb$_2$ (T6)&25.5(2)&20.2(2)&5.40(4)&48.9(1)&0.211(4)&1.03(3)\\
LaAg$_{0.5}$Au$_{0.5}$Sb$_2$ (T2)&25.6(1)&14.8(2)&10.50(8)&49.2(1)&0.415(9)&1.01(3)\\
LaAg$_{0.25}$Au$_{0.75}$Sb$_2$ (T6)&25.7(1)&7.4(1)&17.7(1)&49.28(8)&0.71(1)&1.00(2)\\
\hline \hline
\end{tabular}

\end{table*}

\clearpage

\section{LaAg$_{0.54}$Au$_{0.46}$Sb$_2$ under pressure}

\begin{figure}[h]
\includegraphics[width=10cm]{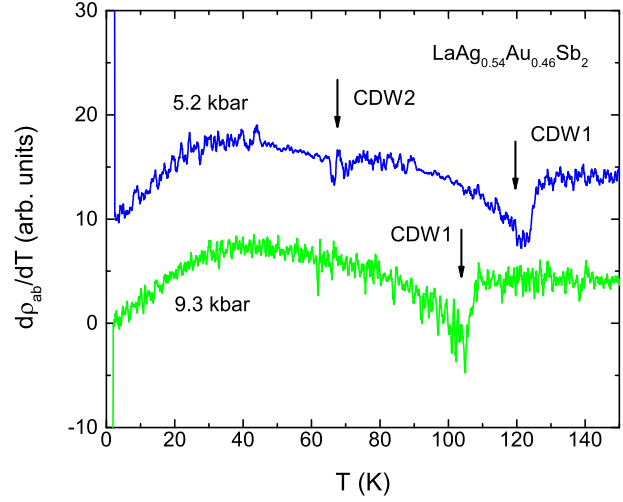}
\caption{(color online) Resistivity derivatives, $d\rho_{ab}/dT$, data at 5.2 kbar and 9.3 kbar  for LaAg$_{0.54}$Au$_{0.46}$Sb$_2$. Arrows point to CDW transition temperatures. The 5.5 kbar data are shifted vertically by 10 for clarity.}
\label{Ader}
\end{figure}

Fig. \ref{Ader} presents the derivatives of the resistivity data taken at 5.2 kbar and 9.3 kbar for LaAg$_{0.54}$Au$_{0.46}$Sb$_2$ sample.

\clearpage
%%%%%%%%%%%%%%%%%%%%%%%%%%%%%%%%%%%%%%%%%%

%=====================================
% References, variant B: internal bibliography
%=====================================

%%%%%%%%%%%%%%%%%%%%%%%%%%%%%%%%%%%%%%%%%%

\end{document}